\documentclass[aps,prb,superscriptaddress,showpacs,twocolumn]{revtex4}
\usepackage{graphicx}
\usepackage{dcolumn}
\usepackage{bm}

\begin{document}
\title{Play building blocks on population distribution of multilevel superconducting flux qubit with quantum interference}
\author{ Xueda Wen}
\affiliation{National Laboratory of Solid State Microstructures and Department of
Physics, Nanjing University, Nanjing 210093, China }
\author{ Yiwen Wang}
\affiliation{National Laboratory of Solid State Microstructures and Department of
Physics, Nanjing University, Nanjing 210093, China }
\author{Yang Yu}
\affiliation{National Laboratory of Solid State Microstructures and Department of
Physics, Nanjing University, Nanjing 210093, China }
\begin{abstract}
Recent experiments on Landau-Zener interference in multilevel superconducting flux qubits
revealed various interesting characteristics, which have been studied theoretically in
our recent work[PRB $\mathbf {79}, 094529$, (2009)] by simply using rate equation method.
In this note we extend this method to the same system but with larger driving amplitude and higher
driving frequency. The results show various anomalous characteristics, some of which have been
observed in a recent experiment.
\end{abstract}

\maketitle
In this note, we show the anomalous characteristics in population distribution of superconducting
flux qubit under driving fields
with large amplitude($A$) and high frequency($\omega$). For a field-driving multilevel flux qubit with diabatic
quantum states $|i,L\rangle$(left well, i=0,1,2...) and $|j,R\rangle$(right well, j=0,1,2,...), \cite{Berns,Oliver}
the Landau-Zener transition rate between $|i,L\rangle$ and $|j,R\rangle$ is:\cite{Wen}
 \begin{equation}\label{Wij}
W_{ij}(\epsilon_{ij},A)=\frac{\Delta_{ij}^{{2}}}{2}\sum_{n}\frac{\Gamma_{2}J^{2}_{n}(x)}{(\epsilon_{ij}-n\omega)^{2}+\Gamma^{2}_{2}},
\label{4}
\end{equation}
where $ \Delta_{ij}$ is the avoided crossing between states $\vert i,L \rangle$ and $\vert j,R \rangle$,
,$\epsilon_{ij}$ is the dc energy detuning from the corresponding avoided crossing $\Delta_{ij}$,
$\Gamma_{2}=1/T_{2}$ is the dephasing rate and  $J_{n}(x)$ are Bessel functions of the first
kind with the argument $x=A/\omega$.

The time evolution of population for state $|i,L\rangle$ can be described by\cite{Wen}
\begin{eqnarray}\label{Rule}
\dot{P}_{i,L}&=&-\sum_{j}W_{ij}P_{i,L}+\sum_{j}W_{ij}P_{j,R}\nonumber\\
&&-\sum_{i'}\Gamma_{i\to i'}P_{i,L}+\sum_{i'}\Gamma_{i'\to i}P_{i',L}\nonumber\\
&&-\sum_{j}\Gamma_{i\to j}P_{i,L}+\sum_{j}\Gamma_{j\to i}P_{j,R},
\end{eqnarray}%
where $\Gamma_{i\to i'}$ is the intrawell relaxation rate from $|i,L\rangle$ to $|i',L\rangle$ ,
and $\Gamma_{i\to j}$ is the interwell relaxation rate from $|i,L\rangle$ to $|j,R\rangle$.
The time evolution of population for state $|j,R\rangle$ can be obtained in the same way.
In the stationary case, we have $\dot{P}_{i,L(R)}=0$. The qubit population distribution can be easily
obtained by
\begin{equation}\label{P}
 P_{L(R)}=\sum_{i(j)}P_{i,L(j,R)}
\end{equation}

Based on Eq.(\ref{Wij}-\ref{P}), we can play building blocks. Eq.(\ref{Wij}) serves as the basic building block,
Eq.(\ref{Rule}) serves as the rule of the game, and Eq.(\ref{P}) shows the total pattern. By adjusting the
frequency $\omega$ and dephasing rate $\Gamma_2$ in Eq.({\ref{Wij}}), we can choose different kinds of building blocks.
Therefore, kinds of interesting patterns can be constructed. As shown in Fig. 1, the diamond patterns observed in
recent experiments\cite{Berns,Oliver} are well obtained based on Eq.(\ref{Wij}-\ref{P}) with experimental parameters.
\begin{figure}
\centering
\includegraphics[width=3.3075in]{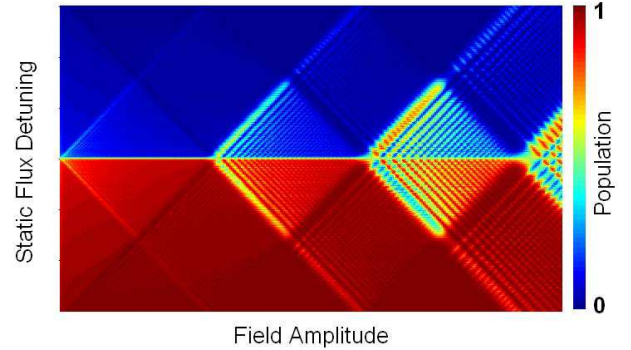}
\caption{Calculated qubit population versus flux detuning and driving amplitude. The parameters are from experiments.\cite{Berns}
The dephasing rate is $\Gamma_2/2\pi=0.5$GHz and the driving frequency is $\omega/2\pi=160$MHz.
}
\end{figure}
By increasing the driving frequency continually, we can obtain more interesting patterns. As shown in Fig. 2,
with the parameters in a recent experiment where the driving frequency is much higher($\omega/2\pi$$\sim$10GHz), we can
obtain various anomalous patterns, some of which have been demonstrated by experiments.\cite{Wang}
\begin{figure}
\centering
\includegraphics[width=5.4375in]{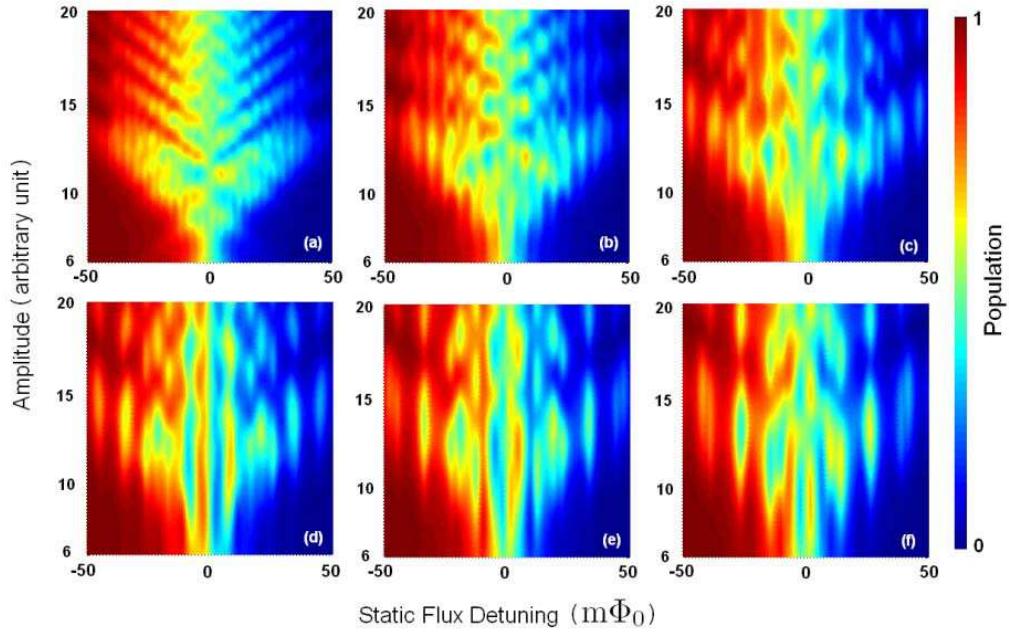}
\caption{Calculated qubit population versus flux detuning and driving amplitude. The simulation parameters are from
experiments.\cite{Wang} The driving frequencies($\omega/2\pi$) are (a)5GHz, (b)8GHz, (c)11GHz, (d)13GHz, (e)15GHz, and
(f)17GHz, respectively. }
\end{figure}

\end{document}